\documentclass[nobibnotes,nofootinbib,aps,prb,twocolumn,floatfix,floats,showpacs,superscriptaddress,amsmath,amsfonts,amssymb]{revtex4-1}

\usepackage{graphicx,xcolor,latexsym}
\usepackage[colorlinks=true, linkcolor=blue, citecolor=blue, urlcolor=blue,
            pdftitle={Excitation of collective modes in a quantum flute}, 
            pdfauthor={Kristinn Torfason}]{hyperref}
\usepackage[all]{hypcap}
\usepackage{ulem} 

\newcommand{\ud}{\mathrm{d}}
\newcommand{\tr}{\operatorname{Tr}}

\def\clap#1{\hbox to 0pt{\hss#1\hss}}

\def\mathrlap{\mathpalette\mathrlapinternal}

\def\mathrlapinternal#1#2{%
\rlap{$\mathsurround=0pt#1{#2}$}}

\newcommand{\brownsolid}{$\vcenter{\hbox{\protect\includegraphics{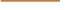}}}$}
\newcommand{\blackdash}{$\vcenter{\hbox{\protect\includegraphics{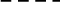}}}$}
\newcommand{\violetdot}{$\vcenter{\hbox{\protect\includegraphics{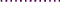}}}$}
\newcommand{\bluedash}{$\vcenter{\hbox{\protect\includegraphics{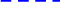}}}$}
\newcommand{\redsolid}{$\vcenter{\hbox{\protect\includegraphics{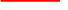}}}$}
\newcommand{\greensolid}{$\vcenter{\hbox{\protect\includegraphics{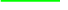}}}$}
\newcommand{\bluecircle}{\protect\includegraphics{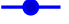}}
\newcommand{\redsquare}{\protect\includegraphics{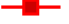}}
\newcommand{\browndiamond}{\protect\includegraphics{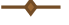}}

\begin{document}


\title{Excitation of collective modes in a quantum flute}


\author{Kristinn Torfason}
\affiliation{School of Science and Engineering, Reykjavik University, Menntavegur 1, IS-101 Reykjavik, Iceland}
\affiliation{Science Institute, University of Iceland, Dunhaga 3, IS-107 Reykjavik, Iceland}

\author{Andrei Manolescu}
\affiliation{School of Science and Engineering, Reykjavik University, Menntavegur 1, IS-101 Reykjavik, Iceland}

\author{Valeriu Molodoveanu}
\affiliation{National Institute of Materials Physics, P. O. Box MG-7, Bucharest-Magurele, Romania}

\author{Vidar Gudmundsson}
\affiliation{Science Institute, University of Iceland, Dunhaga 3, IS-107 Reykjavik, Iceland}


\date{\today}

\begin{abstract}
We use a generalized master equation (GME) formalism to describe the
non-equilibrium time-dependent transport of Coulomb interacting electrons
through a short quantum wire connected to semi-infinite biased leads.
The contact strength between the leads and the wire is modulated by
out-of-phase time-dependent potentials which simulate a turnstile device.
We explore this setup by keeping the contact with one lead at a fixed location at one
end of the wire whereas the contact with the other lead is placed on
various sites along the length of the wire.  We study the propagation
of sinusoidal and rectangular pulses.  We find that the current profiles
in both leads depend not only on the shape of the pulses, but also on
the position of the second contact. The current reflects standing waves
created by the contact potentials, like in a wind musical instrument
(for example a flute), but occurring on the background of the equilibrium
charge distribution.  The number of electrons in our quantum "flute" device
varies between two and three.  We find that for rectangular pulses the
currents in the leads may flow {\it against} the bias for short time intervals,
due to the higher harmonics of the charge response.  The GME is solved numerically 
in small time steps without resorting to the traditional Markov and
rotating wave approximations.  The Coulomb interaction between the
electrons in the sample is included via the exact diagonalization method.
The system (leads plus sample wire) is described by a lattice model.  
\end{abstract}


\maketitle


\section{Introduction}

The control of transient transport properties of open nanodevices
subjected to time-dependent signals is nowadays considered as the main
tool for charge and spin manipulation. Pump-and-probe techniques allow the
indirect measurement of tunneling rates and relaxation times of quantum
dots in the Coulomb blockade \cite{Fujisawa}. Quantum point contacts and
quantum dots submitted to pulses applied only to the input lead generate
specific output currents \cite{Naser1,Naser2,Lai}. Single electrons
pumping through a double quantum dot defined in an InAs nanowire by
periodic modulation of the wire potential has been observed \cite{Fuhrer},
as well as non-adiabatic monoparametric pumping in AlGaAs/GaAs gated
nanowires \cite{Kaestner}.

Modeling such short-time processes is a serious task because even if
the charge dynamics is imposed by the time-dependent driving 
fields, the geometry
of the sample itself and the Coulomb interaction play also an important
role. A well-established approach to time-dependent transport relies on
the non-equilibrium Greens' function (NEGF) formalism, the Coulomb effects
being treated either via density-functional methods \cite{Kurth1} or
within many-body perturbation theory \cite{Myohanen1}. Alternatively,
equation of motion methods were used for studying pumping in finite and
infinite-$U$ Anderson single-level models \cite{Hernandez}.  The numerical
implementation of these formal methods in the interacting case requires
extensive and costly computational work if the sample accommodates more
than few electrons; as a consequence, accurate simulations for systems
having a more complex geometry and/or complex spectral structure are
not easily obtained.

Recently we reported transport calculations for a two-dimensional
parabolic quantum wire in the turnstile setup \cite{NJP}, neglecting
the Coulomb interaction between electrons in the wire.  The latter is
connected to semi-infinite leads seen as particle reservoirs.  Let us
remind here that the turnstile setup was experimentally realized
by Kouwenhoven {\it et al.} \cite{TSP}.  It essentially involves a
time-dependent modulation (pumping) of the tunneling barriers between
the finite sample and drain and source leads, respectively. During the
first half of the pumping cycle the system opens only to the source lead
whereas during the second half of the cycle the drain contact opens.
At certain values of the relevant parameters an integer number of
electrons is transferred across the sample in a complete cycle.  More
complex turnstile pumps have been studied by numerical simulations, like
one-dimensional arrays of junctions \cite{Mizugaki} or two-dimensional
multidot systems \cite{Ikeda}.

In this work we perform a similar study for an {\it interacting}
one-dimensional quantum wire coupled to an input (source) lead at one
end, while the output (drain) lead can be plugged at any point along its
length. Both contacts are modulated by periodic pulses (sinusoidal or
square shaped). Our study is motivated by the possibility to control the
transient currents through the variation of the drain contact. We shall
see in fact that the flexibility of the drain contact allows us to capture
different responses of the sample to local time-dependent perturbations
which can lead to transient currents with specific shapes. In some sense,
our system works like a `quantum flute', this fact being revealed when
analyzing the distribution of charge within the wire. In particular,
we calculate and discuss the deviation of the charge density from the
mean value for each site of the quantum wire and observe the onset of
standing waves.

The effect of the electron-electron interaction is included in the sample
via the exact diagonalization method while the time-dependent transport
is performed within the generalized master equation (GME) formalism as it
is described in Ref.~\onlinecite{PRBC}. The implemented GME formalism
can be used to describe both the initial transient regime immediately
after the coupling of the leads to the sample and the evolution towards a
steady state achieved in the long time limit.  The GME formalism captures
the transient charging of many-body states and Coulomb blockade effects
\cite{PRBC}.

To the best of our knowledge these are the first numerical simulations
of electronic transport through an interacting quantum turnstile
which is not a quantum dot. We emphasize that most of the studies on
pumping in interacting systems are focused on single-level quantum dots
\cite{Splettstoesser,Sela} in the Kondo regime.  Here we consider a
system with spatial extension where the charge distribution plays an
important role in the transport processes.

We discuss for the first time the effect of contacts' location on the
transient currents. More precisely, we show that if the drain lead is
attached to different regions of the quantum wire the currents in both
leads are considerably affected.

The paper is organized as follows: The model and the methodology are
described in \autoref{sec:model}, the numerical results are presented
in \autoref{sec:results}, and the conclusions in \autoref{sec:conclusion}.

\section{The quantum flute model\label{sec:model}}

\begin{figure}
    \includegraphics[width=1.0\linewidth]{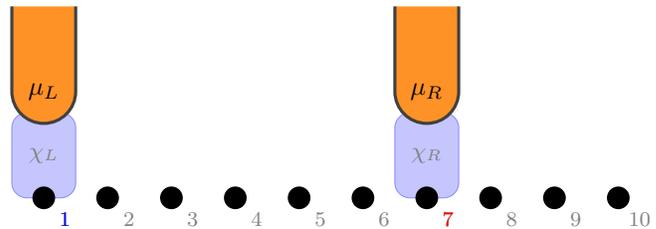}
    \caption{(Color online) A sketch of the system under study. A 1D lattice with 10 sites 
    (``the sample'') is connected to two semi-infinite leads via tunneling.
    The left lead is connected to the left end of lattice, while the position
    of the right lead can be changed. The contacts ($\chi_L$, $\chi_R$) are
    modulated in time.\label{fig:GME-system}}
  \end{figure}

The physical system consists in a sample connected to two leads acting
as particle reservoirs.  We shall adopt a tight-binding description of
the system: the sample is a short quantum wire and the leads are 1D and
semi-infinite.  In this work we consider a sample of 10 sites. This number
optimizes the computational time and the physical phenomenology which we
intend to describe.  A sketch is given in Fig.~\ref{fig:GME-system}.
The left lead (or the source, marked as \textcolor{blue}{$L$}) is
contacted at one end of the sample and the right lead (or the drain,
marked as \textcolor{red}{$R$}) may be contacted on any other site. The
Hamiltonian of the coupled and electrically biased system reads as
  \begin{equation}\label{eq:total_ham}
    H(t) = \sum_{\ell} H_{\ell} + H_S + H_T(t) = H_0 + H_T(t)\, ,
  \end{equation}
where $H_S$ is the Hamiltonian of the isolated sample, including the
electron-electron interaction,
  \begin{equation}\label{eq:sample_ham}
    H_S = \sum_n E_n d_n^\dagger d_n + \frac{1}{2} \sum_{\substack{mn\\m'n'}} 
V_{mn,m'n'} d_m^\dagger d_n^\dagger d_{m'} d_{n'}\, .  
  \end{equation}
The (non-interacting) single-particle basis states have wave functions
$\{\phi_n\}$ and discrete energies $E_n$. $H_{\ell}$, with
${\{\ell\} = (L, R)}$, is the Hamiltonian corresponding to the left
and the right leads. The last term in Eq.~\eqref{eq:total_ham}, $H_T$
describes the time-dependent coupling between the single-particle basis
states of the isolated sample and the states $\{ {\psi}_{q\ell}\}$
of the leads:
  \begin{equation}\label{Htunnel}
    H_T(t)=\sum_{n}\sum_{\ell}\!\int\! \ud q\:\chi_{\ell}(t)(T^{\ell}_{qn}c^{\dagger}_{q\ell}d_n + \text{h.c.}) \,.
  \end{equation}
The function $\chi_{\ell}(t)$ describes the time-dependent switching
of the sample-lead contacts, while $d^{\dagger}_n$ and $ c_{q\ell}$
create/annihilate electrons in the corresponding single-particle
states of the sample or leads, respectively. The coupling coefficient
  \begin{equation}
    T^{\ell}_{qn}=V_0{\psi}^{*}_{q\ell}(0)\phi_n(i_\ell)\, ,
  \end{equation}
involves the two eigenfunctions evaluated at the contact sites
$(0,i_\ell)$, $0$ being the site of the lead $\ell$ and $i_\ell$ the site
in the sample \cite{PRBC}.
In our present calculations we keep the left lead connected to the site
$i_L=1$, while the position of the right lead $i_R$ is varied.
The parameter $V_0$ plays the role of a coupling constant between the 
sample and the leads.
%

We will ignore the Coulomb effects in the leads, where we assume a high
concentration of electrons and thus strong screening and fast particle
rearrangements.  The Coulomb electron-electron interaction is considered
in detail only in the sample, where Coulomb blocking effects may occur.
The matrix elements of the Coulomb potential in Eq.~\eqref{eq:sample_ham}
are given by,
  \begin{equation}
    \!\!\!V_{mn,m'n'} =\! \int\! \ud\vec{x}\ud\vec{x}^{\prime}\, \phi_m^{*}(\vec{x}) 
\phi_{n}^{*}(\vec{x}^{\prime})
\frac{u_C}{|\vec{x}-\vec{x}^{\prime}|} \phi_{m'}(\vec{x}) \phi_{n'}(\vec{x}^{\prime})\, .
  \end{equation}

We calculate the many-electron states (MES) in the sample by
incorporating the Coulomb electron-electron interaction following
the exact diagonalization method, i.~e.\ without any mean field
approximation.  The MES are calculated in the Fock space built on
non-interacting single-particle states \cite{PRBC}.  Since the sample is
open the number of electrons is not fixed, but the Coulomb interaction
conserves the number of electrons.  With 10 lattice sites we obtain 10
single-particle eigenstates and thus ${2^{10}=1024}$ elements in the Fock
space spanned by the occupation numbers.  The Coulomb effects are measured
by the ratio of a characteristic Coulomb energy ${U_C=e^2/(\kappa a)}$
and the hopping energy ${t_s=\hbar^2/(2m_{\mathrm{eff}}a^2)}$. Here $a$
denotes the inter-site distance (the lattice constant of the discretized
system), while $\kappa$ and $m_{\mathrm{eff}}$ are material parameters,
the dielectric constant and the electron effective mass, respectively. In
our calculations we use the {\em relative} strength of the Coulomb
interaction, ${u_C=U_C/t_s}$, which is treated as a free parameter.

%
\section{The transport formalism}

The equation of motion for the our system is the quantum Liouville equation,
  \begin{equation}
   i\hbar \dot{W}(t) = \left[ H(t), W(t) \right]\, ,\qquad W(t < t_0) = \rho_L \rho_R \rho_S\, .  
  \end{equation}
$W(t)$, the statistical operator is the solution of the equation and
completely determines the evolution of the system. At times before $t_0$
the systems are assumed to be isolated and $W(t)$ is simply the product
of the density operator of the sample and the equilibrium distributions
of the leads.

Following the Nakajima-Zwanzig technique \cite{PhysRevB.77.195416}
we define the reduced density operator (RDO), $\rho(t)$, by tracing
out the degrees of freedom of the environment, the leads in our case,
over the statistical operator of the entire system, $W(t)$
  \begin{equation}\label{eq:rdo}
    \rho(t) = \tr_L\tr_R W(t)\, ,\qquad \rho(0) = \rho_S\, .
  \end{equation}
The initial condition corresponds to a decoupled sample and leads when the RDO is
just the statistical operator of the isolated sample $\rho_S$.  For a sufficiently
weak coupling strength ($V_0$) one obtains
the non-Markovian integro-differential master equation for the RDO
  \begin{equation}\begin{split}\label{eq:gme}
    {\dot\rho}(t) &= -\frac{i}{\hbar}[H_S,\rho(t)]\\
                  &-\frac{1}{\hbar^2}\sum_{\ell}\!\int\! \ud q\:\chi_{\ell}(t)
      \Big(\left[{\cal T}_{q\ell},\Omega_{q\ell}(t)\right]+ \text{h.c.} \Big) \, ,
  \end{split}\end{equation}
where the operators $\Omega_{q\ell}$ and $\Pi_{q\ell}$ are defined as
  \begin{equation*}\begin{split}
    \Omega_{q\ell}(t)&=e^{-itH_S}\! \int_{0}^t \!\!\ud s\:\chi_{\ell}(s)\Pi_{q\ell}(s)e^{i(s-t)
    \varepsilon_{q\ell}}e^{itH_S} \, ,\\
    \Pi_{q\ell}(s)&=e^{isH_S}\left ({\cal T}_{q\ell}^{\dagger}\rho(s)(1-f_\ell)-\rho(s)
    {\cal T}_{q\ell}^{\dagger}f_\ell\right )e^{-isH_S} \, ,
  \end{split}\end{equation*}
and $f_\ell$ is the Fermi function of the lead $\ell$ describing the state
of the lead before being coupled to the sample. The operators
${\cal T}_{q\ell}$ and ${\cal T}_{q\ell}^{\dagger}$ describe the
'transitions' between two many-electron states (MES) $|\alpha\rangle$
and $|\beta\rangle$ when one electron enters the sample or leaves it:
  \begin{equation}\label{Tmunuint}
    \left( {\cal T}_{q\ell}\right)_{\alpha\beta}=\sum_n 
    T^\ell_{qn}\langle\alpha|d^{\dagger}_n|\beta\rangle \,.
  \end{equation}
The GME is solved numerically by calculating the matrix elements of the
RDO in the basis of the interacting MES, in small time steps, following a
Crank-Nicolson algorithm. More details of the derivation of the GME can
be found in Ref.~\onlinecite{1367-2630-11-7-073019}.  The calculation of 
the interacting MES is described in Ref.~\onlinecite{PRBC}

Mean values of observables can by obtained by taking the trace of product of 
the corresponding operator and the RDO. The time dependent charge density is 
obtained from the particle-density operator,
${n(x)=\sum_{l,m} \phi_l^*(x) \phi_m(x) d_l^\dagger d_m}$, 
where $\phi_{l,m}(x)$ are single-particle wave functions,
  \begin{equation}
    \langle Q(t, x) \rangle = \sum_{\alpha \beta} \rho_{\alpha \beta}(t) 
\sum_{lm} \phi_l^*(x) \phi_m(x) \langle \beta | d_l^\dagger d_m | \alpha \rangle\, .
  \end{equation}
The total time dependent charge in the sample is found by integrating over $x$ or 
by using the number operator ${\cal N} = \sum_m d_m^\dagger d_m$:
  \begin{equation}\label{eq:obs-chg}
    \langle Q(t) \rangle = e \tr\{\rho {\cal N}\} = e \sum_N N \sum_{\alpha_N} 
\langle \alpha_N | \rho(t) | \alpha_N \rangle\, ,
  \end{equation}
where $\alpha_N$ denotes the (Coulomb interacting) MESs with fixed number of electrons $N$.
Remark that one can also calculate the partial charge accumulated on $N$-particle MESs.

The currents in the system are then found by taking the derivative of 
Eq.~\eqref{eq:obs-chg} with respect to time,
  \begin{equation}\label{eq:obs-crt}
    \langle I(t) \rangle = I_L (t) - I_R(t) = e \sum_N N \sum_{\alpha_N} 
\langle \alpha_N | \dot{\rho}(t) | \alpha_N \rangle\, .
  \end{equation}
The time derivative of the RDO can be substituted by the right-hand side of the 
GME [Eq.~\eqref{eq:gme}] and so it is possible identify the currents in each lead,
  \begin{equation}\begin{split}
    \langle I_\ell(t) \rangle & = -\frac{1}{\hbar^2} \sum_N N \sum_{\alpha_N} 
        \int\! \ud q\, \chi_\ell(t) 
\langle \alpha_N | \left[{\cal T}_{q\ell},\Omega_{q\ell}(t)\right] | 
\alpha_N \rangle \\
& + \text{h.c.}
  \end{split}\end{equation}
We also introduce a \(p\)-indexed period average for the currents (the \(p\)th period 
covers the interval [\(t_{p-1}\),\(t_p\)]):
  \begin{equation}
    i_{p} = \frac{1}{T} \int_{\mathrlap{t_{p-1}}}^{\mathrlap{t_p}}\; \ud t\: I_\ell (t)\, ,
  \end{equation}
which in the periodic phase, i.~e.\ sufficiently long after the initial transient stage,
does not depend on $k$ and on the lead. $T$ is the period of the pulses, and
$Q_p=Tj_p$ is the total charge transferred through the sample within the period $p$.

The switching-functions in Eq.~\eqref{Htunnel} act on the contact regions 
shaded \textcolor{blue!85}{blue} in Fig.~\ref{fig:GME-system} and are used 
to mimic potential barriers with time dependent height.  In the present study we
use two kinds of switching-functions. 
The first switching-function used in the study is a sine function,
  \begin{equation}\label{eq:turnstile_sin}
    \chi_\ell(t) = A\Big\{ 1 + \sin(\omega(t-s) + \phi_\ell) \Big\}\, ,
  \end{equation}
where $A = 0.5$ controls the amplitude and $\omega = 0.105$ the frequency.
The phase shift between the leads is $\pi$, $\phi_L = 0$ and $\phi_R = \pi$.
The last parameter $s = 15$ is used to shift the functions as needed.

The second switching function corresponds to quasi-rectangular pulses, 
and it is made by combining two quasi Fermi functions that are shifted relatively 
to each other,
  \begin{equation}\begin{split}\label{eq:turnstile_pulse}
    \chi_\ell(t) = 1 - \frac{1}{e^{t-\gamma_s^\ell-\delta} + 1}
                 &- \frac{1}{e^{-(t - \gamma_s^\ell) + (T_p^\ell+\delta)} + 1} \, ,\\
    t &\in [0,\, 2T_p^\ell]\, ,
  \end{split}\end{equation}
where $\gamma_s^\ell = \{0,\, T_p^L\}$ defines the phase shift between
the leads ($\ell=L,R$) and $T_p^\ell = 30$ is the pulse length, the
same in the two leads.  The pulses are not built with perfect rectangles for
reasons related to numerical stability. The parameter $\delta$ controls the shape of
the pulse and is fixed at the value $\delta = 10$.  The time unit used
is $\hbar/t_s$.

The time dependent contact functions are graphed at
the bottom of Figs.~\ref{fig:chg-cur-pulse} and~\ref{fig:chg-cur-sin}.
The frequency of the functions in Eq.~\eqref{eq:turnstile_sin} and~\eqref{eq:turnstile_pulse}
were chosen to be similar. The initial values
are $\chi_{L,R} (0) = 0$, i.~e.\ the leads and the 
sample are initially disconnected.
%
%
\section{Results\label{sec:results}}

We will use the relative Coulomb energy ${u_C = 1.0}$.  For a material
like GaAs this value would correspond to a sample length of ${9a \approx
45}$ nm.  Although quite short, this is an experimentally attainable
length. We believe our results are also valid for longer samples, but
the set of our parameters is also restricted by the computational time
spent in solving the GME which grows very fast with the number of MES.
(A typical calculation took several days of CPU.)  The time unit
is $\hbar/t_s \approx 0.029$ picoseconds.  The lead-sample coupling
parameter is also constant, $V_0 = 1.0$ (units of $t_s$).
The chosen parameters are only optimal for the numerical approach, but 
they can possibly be modified to more realistic values if necessary.

\subsection{Energy spectrum}

\begin{figure}
    \includegraphics{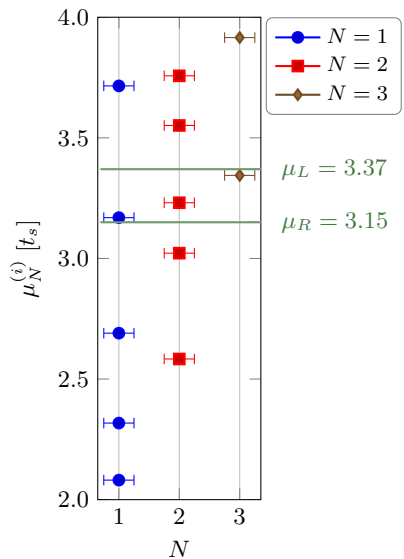}
    \caption{\label{fig:mu-diag}(Color online)
             The energy diagram.
             The \textcolor{blue}{blue} circles (\bluecircle) correspond to single-particle
             states, the \textcolor{red}{red} squares (\redsquare) to two-particle states and the
             \textcolor{brown!60!black}{brown} diamonds (\browndiamond) to three-particle states.
             The \textcolor{green}{green} solid line (\greensolid) is the bias window
             $\Delta\mu = \mu_L - \mu_R = 3.20-2.98 = 0.22$.
             The bias window includes the three particle ground state, but also excited
             one and two-particle states. So the expected number of electrons in
             the steady state is slightly below three.}
  \end{figure}

The MESs of the sample are characterized by the chemical potentials
${\mu_N^{(i)}:={\cal E}_N^{(i)}-{\cal E}_{N-1}^{(0)}}$, where ${\cal
E}_{N}^{(i)}$ is the energy of the interacting MBS number $i$ containing
$N$ particles, $i=0$ indicating to the ground state and $i>0$ the excited
states.  In Fig.~\ref{fig:mu-diag} we show the chemical potential diagram
for our system.  The strength of the Coulomb interaction is $u_C=1$.
For the single-particle states ($N=1$) the chemical potentials are in
fact the single-particle energies.  The effect of the Coulomb interaction
is clearly visible for $N>2$.  For example the lowest chemical potential
for $N=2$, $\mu_2^{(0)} \approx 2.58$, whereas in the absence of Coulomb
interaction it is equal to  $\mu_1^{(1)} \approx 2.32$.

We select the bias window ${\Delta\mu=\mu_L-\mu_R}$ such that it includes
the ground state with $N=3$ electrons, $\mu_L=3.20$ and $\mu_R=2.98$.
The bias window includes also excited single- and two-particle states.
The single-particle state is well below the top of the bias and so the
population of this state will be very small, and the number of electrons
in the sample is expected to be somewhere between 2 and 3.  Therefore we
also expect only two- and three-particle states to be involved in the
transport of electrons through the sample.\cite{PRBC}

\subsection{Sinusoidal pulses\label{sec:results-sine}}

\begin{figure}
    \includegraphics[width=1.0\linewidth]{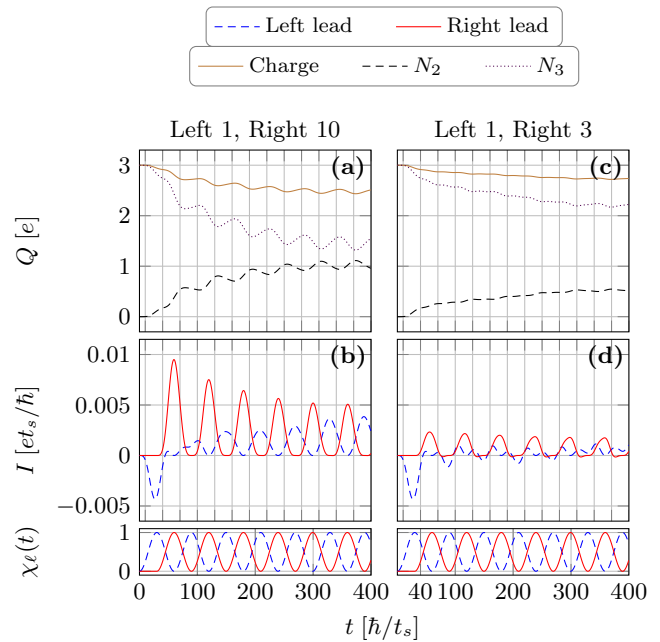}
    \caption{\label{fig:chg-cur-sin}(Color online) Charge and current for
             $\mu_L = 3.37$, $\mu_R = 3.15$, $u_C = 1.0$ and
             $\chi_\ell(t) \propto \sin(\omega t)$.
             Total charge \textcolor{brown}{brown} solid line (\brownsolid),
             charge for two particle states \textcolor{black}{black} dashed (\blackdash),
             for three particle states \textcolor{violet!60!black}{violet} dotted (\violetdot).
             Current for the left lead \textcolor{blue}{blue} dashed (\bluedash),
             for the right lead \textcolor{red}{red} solid (\redsolid).
             We consider two locations of the right lead.
             \textbf{(a)} Charge, left lead 1, right lead 10.
             \textbf{(b)} Current, left lead 1, right lead 10.
             \textbf{(c)} Charge, left lead 1, right lead 3.
             \textbf{(d)} Current, left lead 1, right lead 3.}
  \end{figure}

We begin the time-dependent calculations with $N=3$ electrons in the
sample, initially assumed in the ground state.  This is done by initializing
the diagonal density-matrix element of the sample corresponding to this state
to one and all the other matrix elements to zero.
The left lead ($L$) is
permanently in contact with the left end of the sample, i.~e.\ site 1.
The right lead ($R$) is placed on various other sites, as indicated in  
Fig.~\ref{fig:GME-system}.  The time evolution is then followed in short 
time steps, by using the contact functions $\chi_{L,R}(t)$.  The charge
accumulated in the sample and the currents in the two leads are calculated
at each time step.  In Fig.~\ref{fig:chg-cur-sin} we show the results
with sinusoidal pulses, corresponding to Eq.~\eqref{eq:turnstile_sin},
and with two different placements of the $R$ lead: on sites 10 and 3.

We first observe the charge in the sample and its time evolution shown in
the upper panels of Fig.~\ref{fig:chg-cur-sin}. After the contacts begin
to operate the initial charge of $N=3$ electrons changes in time. Part of
the charge flows into the leads, depending on which one is accessible,
and the average charge drops, until a periodic regime is established.
The lower panels of the figure show the contact functions.
In Fig.~{\ref{fig:chg-cur-sin}\hyperref[fig:chg-cur-sin]{(a)}} the right contact
is placed at site 10.
The charge in the sample has maxima and minima at the time points when
the contact functions are equal. In between these time points, the charge
increases when the left contact opens further and the right one closes,
and decreases otherwise.  The population of the three-particle and
two-particle states are also shown (the population of the single-particle
state being negligible) and they oscillate in antiphase, i.~e.\  
the gain of one is partly compensated by the loss of the other one.

The currents in the leads are shown
in~{\ref{fig:chg-cur-sin}\hyperref[fig:chg-cur-sin]{(b)}}, and they have
similar shape as the contact functions, except in the initial transient phase,
before the periodic regime is stabilized.  In the first cycle the current
in the left lead is initially negative.  The sign rule is that positive
currents correspond to charge flow from the left to the right lead, and
negative currents correspond to the opposite direction.  The initial
negative current in the left lead indicates initial charge flow from
the sample into that lead during the first cycle as long as the contact 
to the right lead is closed.  The main impression of these results is that
the periodic regime qualitatively corresponds to a linear response of
the charge and currents to the contact strength.

The situation may change when the right contact is placed on another
site, for example on site 3, as shown in~{\ref{fig:chg-cur-sin}\hyperref[fig:chg-cur-sin]{(c-d)}}.
In this case the oscillations of the charge and currents are weaker.
The current in the left lead is no longer sinus-like.  This shows now
a non-linear behavior of the charge response to the same pulses as
before. Some sort of standing waves are created in the sample and the
right contact creates a local perturbation of the charge fluctuations
at that point.  Negative currents in the left lead may occur now during
more pulse cycles as before.  This is somewhat surprising, since such
currents, although very small, are actually driven against the bias.
Let's mention that in the absence of a bias ($\Delta \mu =0$) the
currents in both leads oscillate between positive and negative values,
but with zero average, such that no real pumping effect is obtained in this
setup, irrespectively of the placement of the leads (not shown).

The currents in the leads reflect the charging or discharging of the
sample, but these are actually complex processes, because different
states may be occupied with different time constants, related to the
tunneling matrix elements, and thus the charging and the currents may
have short-time fluctuations. The fine structure of the currents is thus
a complicated issue, which will be discussed further. 

Before moving to that, we should remark that our charging and
current curves are smooth, whereas in a real experiment the electron
tunneling may look like a stochastic (discontinuous) process. Of course in this
work we are not describing the electron dynamics at that level.
The method of GME gives only expected
values of charge and currents in the quantum mechanical sense,
Eqs.~(\ref{eq:obs-chg}) and (\ref{eq:obs-crt}).  In the example of Fig.~\ref{fig:chg-cur-sin}
the pulse frequency is comparable to the average tunneling rate, which is 
proportional to the square of the coupling parameter $V_0^2$. But the currents are 
also smooth because of 
the sinusoidal pulses, and they will look different for rectangular pulses 
discussed in \autoref{sec:results}~\hyperref[sec:results-pulse]{D}.

\subsection{Charge distribution in the sample}

\begin{figure}
    \includegraphics[]{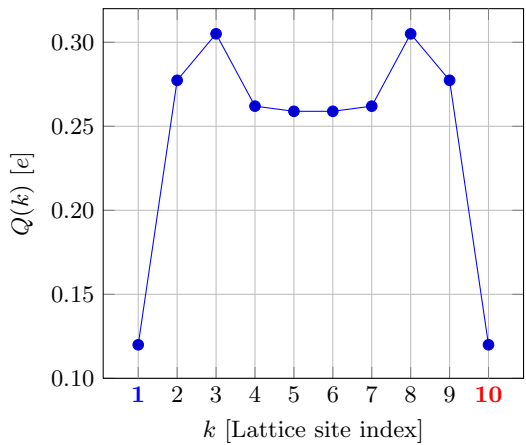}
    \caption{\label{fig:average_charge}(Color online) The charge distribution along 
    the sample averaged in time over the whole period $t = 310$ to $t = 370$, when the periodic
    regime is established. The right contact is placed on site 10 (marked in red). 
    The main parameters are $\mu_L = 3.37$, $\mu_R = 3.15$ and $u_C = 1.0$.
}
  \end{figure}

The charge distribution inside the sample is shown in the
Fig.~\ref{fig:average_charge} and it is far from homogeneous.  The charge
is averaged in time over an entire period of the contact functions when
the system is in a periodic regime. In the case shown the right contact
is placed on site 10.  The charge distribution does not qualitatively
change for other placements of the right contact (not shown).  The distribution
is symmetric along the sample, in spite of the presence of the bias window, 
which shows that the contacts between the sample and the leads are actually 
weak in our case.  We can say that the charge distribution follows the geometrical 
extend of those single-particle states that contribute to the active two-
and three-particle MBS.

\begin{figure}
    \includegraphics[width=1.0\linewidth]{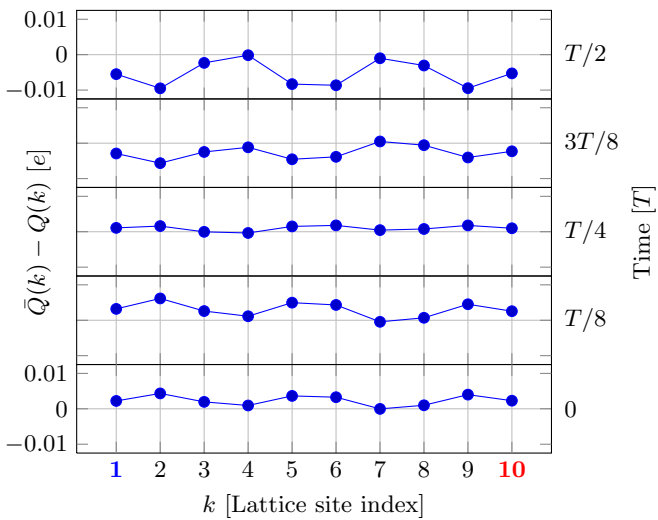}
    \caption{\label{fig:charge_wave1}(Color online) Snapshots of the deviation of charge from the average
             over the half-period $t = 340$ to $t = 370$.
             For $\omega=0.105$ and $\chi_\ell(t) \propto \sin(\omega t)$ with the right contact at site 10.
             A video showing the time dependent charge oscillations and the currents in the leads can be accessed 
             online on arXiv in ancillary files.}
  \end{figure}

\begin{figure}
    \includegraphics[width=1.0\linewidth]{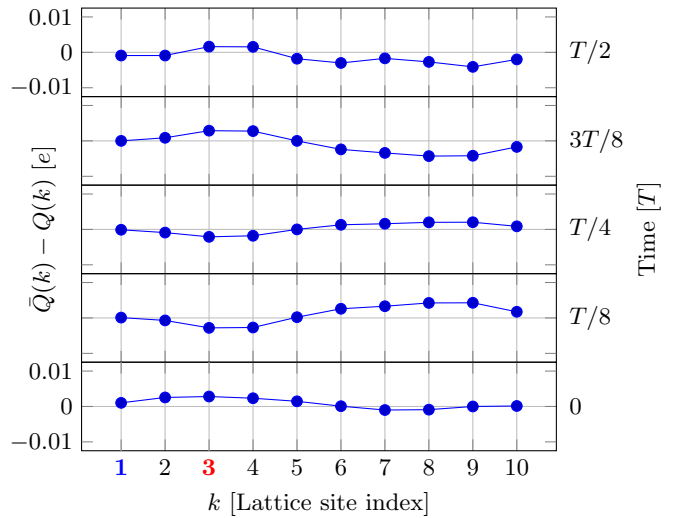}
    \caption{\label{fig:charge_wave2}(Color online) Snapshots of the deviation of charge from the average
             over the half-period $t = 340$ to $t = 370$.
             For $\omega=0.105$ and $\chi_\ell(t) \propto \sin(\omega t)$ with the right contact at site 3.}
  \end{figure}

Next, in Figs.~\ref{fig:charge_wave1} and~\ref{fig:charge_wave2}, we
show the deviation of the charge density from the mean value, on each
lattice site, for selected time moments during half a cycle. For the
other half-cycle the reverse motion occurs.   The two placements of
the right lead ate again selected at sites 10 and 3.  Standing
waves are clearly seen.  For the contact configuration L1R10
(Fig.~\ref{fig:charge_wave1}) the standing-wave pattern shows something between
two and three wavelengths. Nodes and antinodes can be 
distinguished, and also a global up and down motion mode seems to
occur.  But it is clear that the amplitude of the charge oscillations
at the contact sites is quite large.  

When the right contact is on
site 3 (Fig.~\ref{fig:charge_wave2}) only about two wavelengths may
be seen, at least for ${t<3T/8}$, but also the charge seems to oscillate
with different amplitudes at the two contacts, larger at the right lead
than at the left lead.  This explains the amplitudes of the currents
seen in Figs.~{\ref{fig:chg-cur-sin}\hyperref[fig:chg-cur-sin]{(b)}}
and~\hyperref[fig:chg-cur-sin]{(d)}.  In other words the
amplitude of the currents in the leads is related to the amplitude of
the charge fluctuation at the contact.  In addition, the finer structure
of the current pulses reflects the existence of higher harmonics in
the charge oscillations. This is suggested by the top part of
Fig.~\ref{fig:charge_wave2}, where we see a higher order pattern at ${t=T/2}$
than at $t=0$.

For a better visual description of the time dependent charge oscillations
we prepared a number of video files which can be accessed online, see
ancillary files
({\tt\mbox{Chg-osc-L1R2-sine.mp4}}, {\tt\mbox{Chg-osc-L1R8-sine.mp4}} and {\tt\mbox{Chg-osc-L1R10-sine.mp4}})
on arXiv. The time dependent charge fluctuations and
the current profiles are shown in the videos for three placements of the
right leads, R2, R8, and R10, respectively.  The behavior of the system
shows the same symmetry as the charge distribution: the current pulses
are similar when the right contact is placed on the right or on the left
of the center of the sample, but at the same distance.  For example R3
and R8, or R4 and R7, etc., are similar.  We are seeing simple collective
oscillations onset by the bias field. The counteracting (restoring) force
comes from the sample-lead boundaries and from the Coulomb interaction.

\subsection{Rectangular pulses\label{sec:results-pulse}}

\begin{figure*}
    \includegraphics[width=1.0\linewidth]{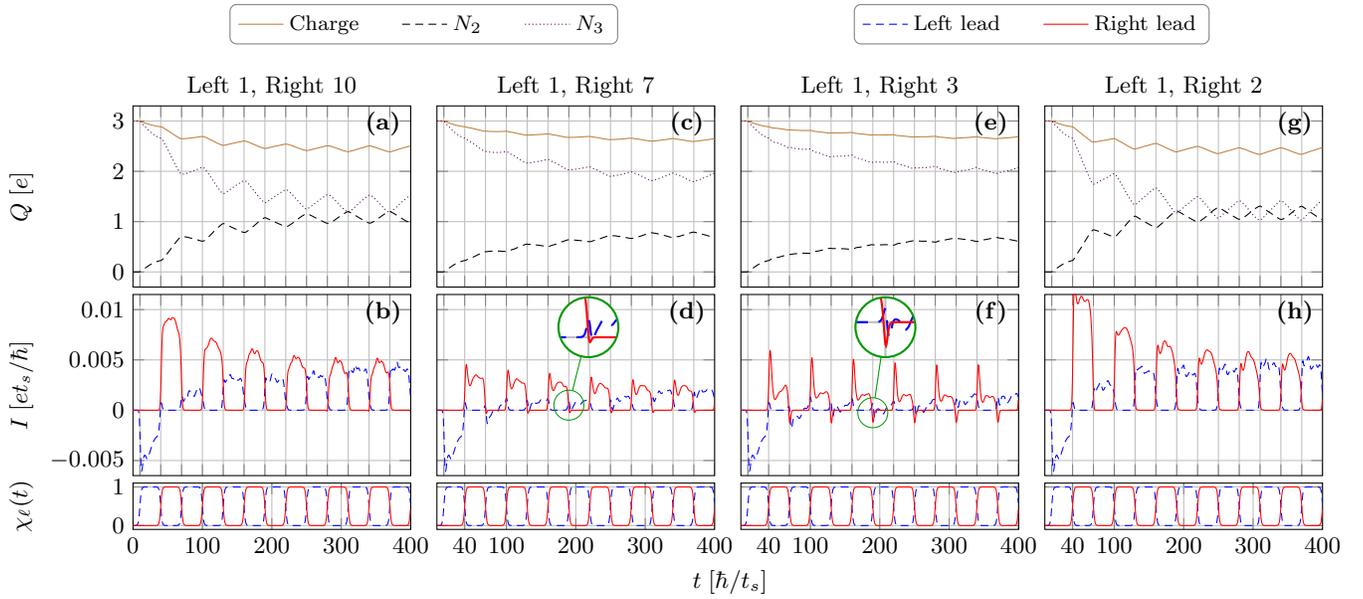}
    \caption{\label{fig:chg-cur-pulse}(Color online) Charge and current for the pulses.
             Total charge \textcolor{brown}{brown} solid line (\brownsolid),
             Charge for two particle states \textcolor{black}{black} dashed (\blackdash),
             Charge for three particle states \textcolor{violet!60!black}{violet} dotted (\violetdot),
             Current for the left lead \textcolor{blue}{blue} dashed (\bluedash),
             Current for the right lead \textcolor{red}{red} solid (\redsolid).
             $\mu_L = 3.37$, $\mu_R = 3.15$, $u_C = 1.0$, $\chi_\ell(t)$ pulses.
             \textbf{(a)} Charge left lead 1, right lead 10.
             \textbf{(b)} Current left lead 1, right lead 10.
             \textbf{(c)} Charge left lead 1, right lead 7.
             \textbf{(d)} Current left lead 1, right lead 7.
             \textbf{(e)} Charge left lead 1, right lead 3.
             \textbf{(f)} Current left lead 1, right lead 3.
             \textbf{(g)} Charge left lead 1, right lead 2.
             \textbf{(h)} Current left lead 1, right lead 2.}
  \end{figure*}

Next we consider the more complex case of the quasi-rectangular contact functions, 
i.~e.\ described by Eq.~\eqref{eq:turnstile_pulse}.  Again the left lead is
permanently in contact with the left end of the sample, and the right lead 
is placed on other sites.  The representative results are shown in 
Fig.~\ref{fig:chg-cur-pulse}, now with four different placements of the $R$ 
lead: on sites 10, 7, 3, and 2, respectively. The charge evolution in time is 
not visibly different from the previous case of the harmonic pulses.
3 electrons, depending on the placement of the $R$ lead.  The populations
of the two-particle and three-particle states have opposite variations in
time: the gain of one is partly compensated by the loss of the other one.
It appears that the weakest charge oscillations occur when the drain lead ($R$) is 
coupled to the site 3, like before for the sine pulses. 
The calculated currents, shown in the middle panels of Fig.~\ref{fig:chg-cur-pulse},
look now more complex than those obtained for
harmonic pulses.  Obviously the rectangular pulses activate more higher
harmonics of the charge oscillations.  Still, in those cases when the
amplitude of the charge oscillations is (relatively) large, i.~e.\ when
the $R$ contact is on the sites 10 and 2, the current oscillations have
almost a rectangular shape, qualitatively reproducing the shape of the
contact functions.  After the initial irregular transient oscillations
the currents become positive in both leads, describing charge propagation
in the direction imposed by the the bias, i.~e.\ from left lead to right.

The current profile is qualitatively different in the other cases,
when the charge oscillations have small amplitude. Sharp and multiple
oscillations are now visible in the currents, produced by higher harmonics
of the charge motion, but invisible in the charge diagrams. Also, even
negative currents can be seen, now in both leads, although small and only
for short times, indicating again charge propagation against the bias. But
now such negative currents, although weak, do not vanish in the long time
limit, when the system approaches a periodic evolution in time.  We thus
see that the placement of the right contact is qualitatively important
for the current profiles for a finite bias window.  
Increasing the bias window the negative currents may survive in one lead
only, and further they disappear, and the current pulses take more and
more the shape of the contact functions.

As for the sine pulses, the current profiles obtained for rectangular pulses
obey the symmetry of the charge distribution. This means we obtain similar results
for the right contact at sites 10 or 1, 9 or 2, 8 or 3, 4 or 7, and 5 or 6.  
Each case is shown in Fig.~\ref{fig:chg-cur-pulse} once, except the later two, 
R5 and R6, which actually look qualitatively similar to R10, only slightly sharper.

\begin{figure}
    \includegraphics[width=1.0\linewidth]{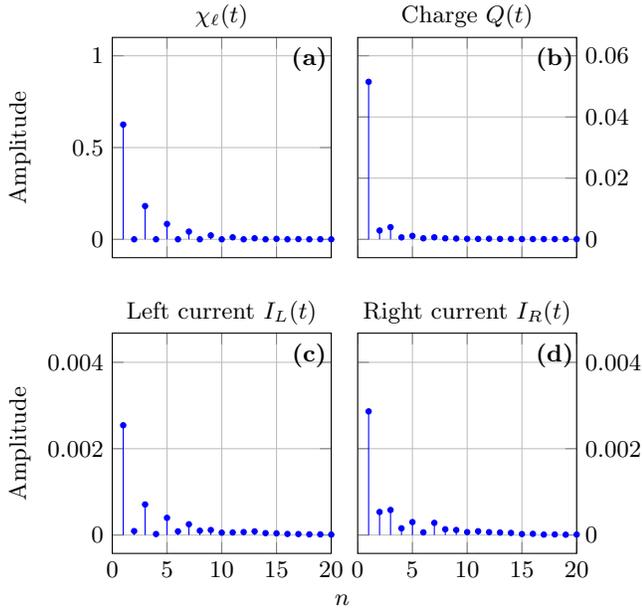}
    \caption{\label{Fourier_pulse_L1R10}(Color online)
             $\mu_L = 3.37$, $\mu_R = 3.15$, $u_C = 1.0$ and $\chi_\ell(t)$
             a rectangular pulse with the right lead connected at site 10.
             Fourier analysis of,
             \textbf{(a)} the switching-function $\chi_{\ell}(t)$,
             \textbf{(b)} the total charge,
             \textbf{(c)} the left current,
             \textbf{(d)} the right current.}
  \end{figure}

\begin{figure}
    \includegraphics[width=1.0\linewidth]{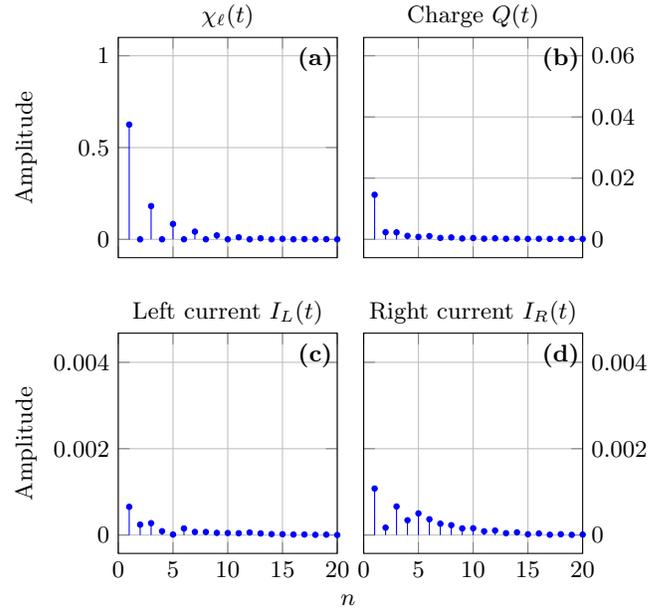}
    \caption{\label{Fourier_pulse_L1R3}(Color online)
             ${\mu_L = 3.37}$, ${\mu_R = 3.15}$, ${u_C = 1.0}$ and $\chi_\ell(t)$
             a rectangular pulse with the right lead connected at site 3.
             Fourier analysis of,
             \textbf{(a)} the switching-function $\chi_{\ell}(t)$,
             \textbf{(b)} the total charge,
             \textbf{(c)} the left current,
             \textbf{(d)} right current.}
  \end{figure}

In order to observe better the effects of higher harmonics of the charge oscillations
we show in Fig.~\ref{Fourier_pulse_L1R10} and Fig.~\ref{Fourier_pulse_L1R3} the Fourier analysis
for two selected contact configurations of rectangular pulses, L1R10 and L1R3 respectively.
Figs.~{\ref{Fourier_pulse_L1R10}\hyperref[Fourier_pulse_L1R10]{(a)}}
and~{\ref{Fourier_pulse_L1R3}\hyperref[Fourier_pulse_L1R3]{(a)}}
show the Fourier components of the switching-function $\chi_\ell(t)$ in Eq.~\eqref{eq:turnstile_pulse}.
The current approximately follows the pulse shape in the left lead when contacts are placed at L1R10
(Fig.~{\ref{Fourier_pulse_L1R10}\hyperref[Fourier_pulse_L1R10]{(c)}}). This behavior is not seen in
the right lead (Fig.~{\ref{Fourier_pulse_L1R10}\hyperref[Fourier_pulse_L1R10]{(d)}}) or
when the contacts are placed at L1R3 (Fig.~{\ref{Fourier_pulse_L1R3}\hyperref[Fourier_pulse_L1R3]{(c-d)}}).
The Fourier components of the total charge are shown in
Figs.~{\ref{Fourier_pulse_L1R10}\hyperref[Fourier_pulse_L1R10]{(b)}}
and~{\ref{Fourier_pulse_L1R3}\hyperref[Fourier_pulse_L1R3]{(b)}}. The first harmonic is the most dominant 
for the configuration L1R10, but for L1R3 (see Fig.~\ref{Fourier_pulse_L1R3}) the cumulative contribution of 
higher harmonics is almost comparable to the main component.
%

\subsection{Charge propagation}

\begin{figure}
    \includegraphics[width=1.0\linewidth]{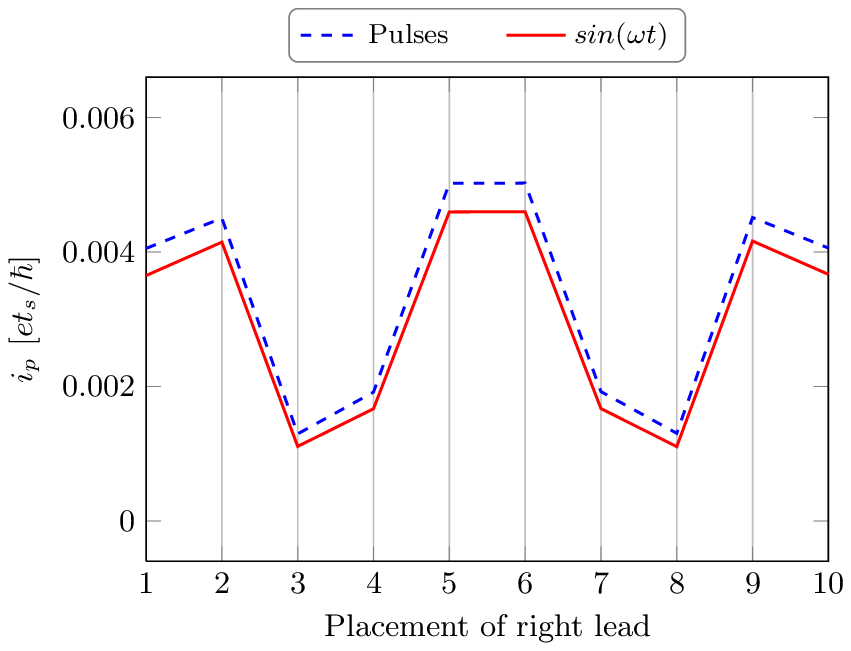}
    \caption{\label{transf_charge}(Color online) Average current (Charge transferred)
             vs placement of the right lead.
             ${\mu_L = 3.37}$, ${\mu_R = 3.15}$, ${u_C = 1.0}$.
             The \textcolor{blue}{blue} dashed (\bluedash) line for $\chi_\ell(t)$ as rectangular pulses and
             \textcolor{red}{red} solid (\redsolid) for $\chi_\ell(t) \propto \sin(\omega t)$.}
  \end{figure}

Next we discuss the amount of charge which propagates through the sample during
one period by our simulated turnstile pump, $Q_k=TI_k$, where $k$ is the right contact
site and $I_k$ the current averaged over one pulse period, from $t \approx 280$ to $t \approx 340$. 
This is shown in Fig.~\ref{transf_charge} where we compare the results for the two type
of pulses.  Interestingly, the pumped charge increases for rectangular pulses.  This
happens especially for those contact locations which produce quasi-rectangular currents,
1, 2, 5, 6, 8, 10.  This can be attributed to higher harmonics of the charge oscillations.

\begin{figure}
    \includegraphics[width=1.0\linewidth]{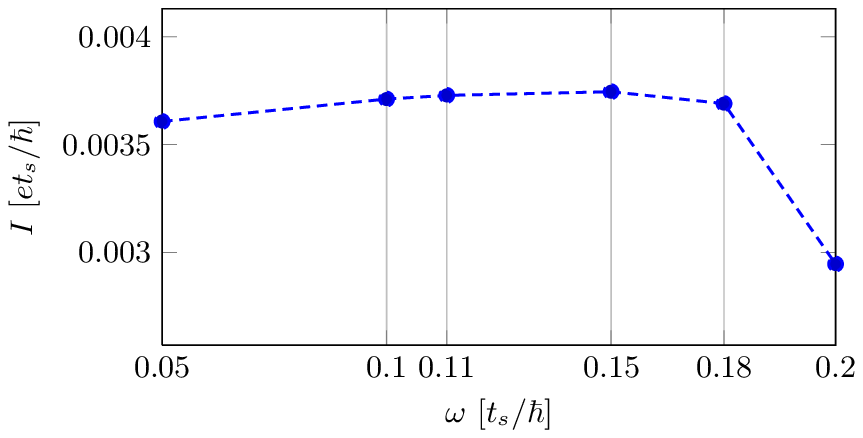}
    \caption{\label{fig:omega_charge}(Color online) Average current vs
    $\omega$. Left lead 1, right lead 10. ${V_0 = 1.0}$, ${u_C = 1.0}$,
             initial state three particle ground state.}
  \end{figure}

Until now all results have been obtained for a fixed frequency of
the contact pulses $T=60$ time units, which correspond to an angular
frequency ${\omega=2\pi/T \approx 0.105}$.  In Fig.~\ref{fig:omega_charge} we show
the average current (which gives the transferred charge) for a variable
frequency, in the interval 0.05-0.2.  The curve is smooth, but the maximum
current is obtained for a frequency between 0.11-0.15.
The frequency corresponding to the maximum
current can be related to the energy spectrum, Fig.~\ref{fig:mu-diag}.
The bias window includes the two-particle and three-particle states
with chemical potentials ${\mu_2^{(2)}=3.231}$ and ${\mu_3^{(0)}=3.344}$.
The difference between these values, 0.113, is in the frequency interval
containing the maximum current, which in fact describes a resonance
between two- and three-particle states. The resonance is broad because
of the effect of the contacts in the electron states in the sample.
The strength of the lead-sample coupling $| V_0T_{qn}^l |^2$ is of the
order 0.02 in our calculations.
At larger frequencies (larger than 0.18) the pulses become to fast so the
charge in the sample cannot follow the imposed time evolution, and so the
current drops.

  \begin{figure}
    \includegraphics[width=1.0\linewidth]{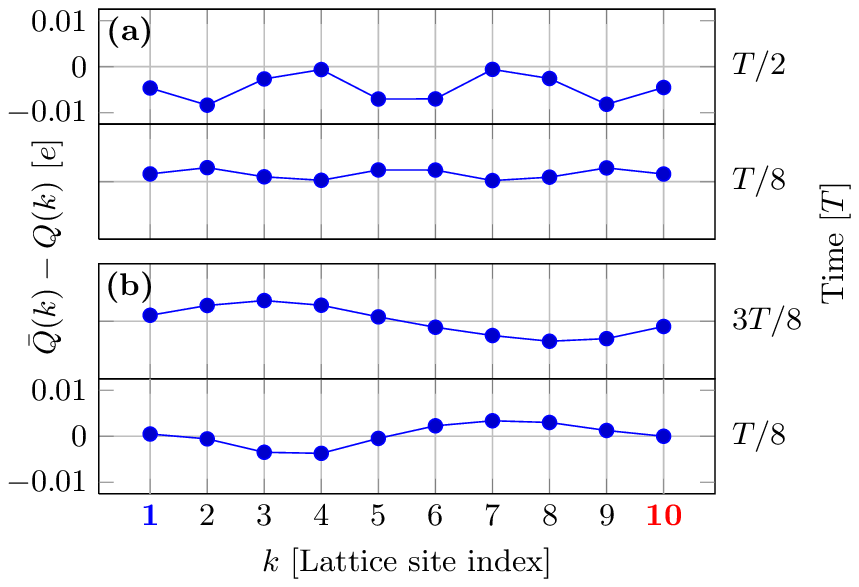}
    \caption{\label{fig:charge_wave3}(Color online) Snapshots of the deviation of charge from the average
             over the half-period ${t = 340}$ to ${t = 370}$.
             For ${\chi_\ell(t) \propto \sin(\omega t)}$ with the right contact at site 10,
             \textbf{(a)}~${\omega = 0.05}$,
             \textbf{(b)}~${\omega = 0.20}$.}
  \end{figure}
%


Finally, in Fig.~\ref{fig:charge_wave3} we show again
the deviation of the charge density from the mean value,
i.~e.\ the standing waves, for two frequencies different
from the frequency used in
\autoref{sec:results}~\hyperref[sec:results-sine]{B}-\hyperref[sec:results-pulse]{D}
($\omega=0.105$). In
Fig.~{\ref{fig:charge_wave3}\hyperref[fig:charge_wave3]{(a)}} the
frequency of the switching-function in Eq.~\ref{eq:turnstile_sin}
has been lowered to ${\omega = 0.05}$ while in
Fig.~{\ref{fig:charge_wave3}\hyperref[fig:charge_wave3]{(a)}} it has
been raised to ${\omega = 0.20}$. The standing-waves observed are
qualitatively similar to those seen in Figs.~\ref{fig:charge_wave1}
for ${\omega = 0.05}$, although of a slightly smaller amplitude, for
example at $T/8$, which is consistent with Fig. \ref{fig:omega_charge}. 
For ${\omega = 0.20}$ we find the higher modes attenuated and apparently
a longer wavelength.  However, a systematic investigation of the dispersion
of the standing waves is not possible at this stage.

\section{Concluding remarks\label{sec:conclusion}}


We have simulated the time dependent transport through a one-dimensional
interacting finite quantum wire attached to two leads, described by
a lattice model. Out-of-phase time periodic signals are applied at the
contact sites generating a turnstile operation. The calculations are
performed by solving the generalized master equation of the reduced
density operator in the Fock space of the many-body states of the
electrons in the sample, which are calculated via exact diagonalization.
The time periodic contacts generate standing waves of charge along the
length of the wire with finer details depending on the location of the
contacts. The amplitude of the currents depends on the contact site
not only through the tunneling constants but also through the amplitude
of the charge fluctuations at the given contact. The current profile
in the leads  is determined by the charge fluctuations in the sample,
which depend on the entire standing wave pattern.  We have called this
model a quantum flute.


We emphasize that the excitation of these collective oscillations
are obtained from a fully quantum treatment of an open few-electron
system.  Collective oscillations have been studied in quantum dots
with few electrons by mapping out the oscillator strength of exact MESs
\cite{col1} or a by more traditional linear response approach \cite{col2}.
The evolution to systems with a higher number of electrons has also been
studied \cite{col3}.  In the present work the excitation method is of
a different type and it may be used in transport measurements.


The charge oscillations occur both due to the finite length of the
sample and due to the Coulomb interaction.  Both the boundaries of
the sample and the Coulomb repulsion create the restoring (``elastic'')
forces.  The Coulomb interaction has also an important role in the
charge distribution, through the well known blocking and correlation
effects, and therefore the inclusion of the Coulomb effects is necessary
for a consistent description.  The relative strength of the Coulomb
interaction $u_C$ may however depends on the material constants.  It is
quite difficult to compare the results without and with the Coulomb
interaction included for the same set of parameters (i.~e.\ bias window,
coupling constants). If Coulomb effects are neglected ($u_C=0$) the
whole energy spectrum changes and new states are present within the bias
window. Then the chemical potentials in the leads have to be shifted
accordingly in order to capture in the bias window states with similar
number of electrons as in the interacting case. This means one cannot
compare the two situations just by changing only one parameter.


Our calculations are performed for a relatively weak coupling of the 
sample to leads. In this case the perturbation of the sample states due to 
the leads is minimal, and so the results can still be interpreted in terms of the
states of the sample itself.  Increasing the coupling strength the finer 
details of the current profile determined by the sample may be washed out.


The standing waves are obviously not possible in quantum dots where, to our 
knowledge, most of the experimental and theoretical work on turnstile pumping
has been done.  Therefore the spatial extension of the sample is, in our opinion,
a novel element in this topic.  The short time scale of the oscillations, in the 
picosecond domain, is related to the energy gaps between the quantum states.
This domain is attainable by the present experimental technology.


In this work we completely ignored the Coulomb interaction in the leads.
Due to computational limitations we studied a short nanowire which
generated a short time scale. And also a relatively narrow bias window,
which together with the weak sample-leads coupling yielded low currents.
But even though the currents calculated in these examples are small
and we are limited here only to the qualitative effects, the predicted
results may be seen in future experiments. For example in a setup with an
array of finger electrodes placed on top of a single wire where currents
can be driven by any pair of contacts \cite{Frielinghaus,Blomers}.
Another possibility is to change the location of one contact,
like in our simulations, using a scanning tunneling or a conductive atomic force microscope
\cite{Zhou}. Consequently one can expect that a suitable placement of
the source and drain leads along the sample would be a way to deliver
modulated output currents with a desired shape and period.  The time
dependent charge propagation along transmission lines of a quantum
mechanical nature may be a future direction in the field of nanophysics.

\begin{acknowledgments}
This work was financially supported by the Icelandic Research Fund. V.M. was also 
supported by the grant of the Romanian National Authority for Scientific
Research, CNCS – UEFISCDI, project number PN-II-ID-PCE-2011-3-0091.
\end{acknowledgments}



%

\end{document}